\documentclass[twocolumn]{aastex631}

\emergencystretch=\maxdimen
\begin{document}

\title{Distinguishing the Demographics of Compact Binaries with Merger Entropy Index}

\author{Siyuan Chen}
\affiliation{Department of Physics and Astronomy, Vanderbilt University \\
2301 Vanderbilt Place, Nashville, TN, 37235, USA}

\author{Karan Jani}
\affiliation{Department of Physics and Astronomy, Vanderbilt University \\
2301 Vanderbilt Place, Nashville, TN, 37235, USA}

\begin{abstract}
The coalescence of binary black holes and neutron stars increases the entropy in the universe. The release of entropy from the inspiral stage to the merger depends primarily on the mass and spin vectors of the compact binary. In this study, we report a novel application of entropy to study the demographics of the compact binaries reported by the LIGO-Virgo-KAGRA (LVK) Collaboration. We compute an astrophysical distribution of the Merger Entropy Index ($\mathcal{I}_\mathrm{BBH}$) - a mass-independent measure of the efficiency of entropy transfer for black hole binaries - for all the events reported in the LVK Gravitational-Wave Transient Catalogs. We derive $\mathcal{I}_\mathrm{BBH}$ for six astrophysically motivated population models describing dynamical and isolated formation channels. We find that $\mathcal{I}_\mathrm{BBH}$ offers a new criterion to probe the formation channels of LVK events with compact objects in the upper $(\gtrsim 60~M_\odot)$ and lower ($\lesssim 5~M_\odot$) mass-gaps. For GW190521, an event with both objects in the upper mass gap, $\mathcal{I}_\mathrm{BBH}$ distribution strongly favors second-generation mergers. For GW230529, a new event with the primary object in the lower mass gap, we note that $\mathcal{I}_\mathrm{BBH}$ mildly favors it with neutron star - black holes events. Our work provides a new framework to study the underlying demographics of compact binaries in the data-rich era of gravitational-wave astronomy.
\end{abstract}

\section{Introduction} \label{sec:intro}
All the gravitational-wave (GW) events observed by the LIGO-Virgo-KAGRA (LVK) Collaboration till date are associated with the coalescence of compact binaries \citep{LIGOScientific:2016vlm}. During the first three observing runs from 2015 to 2020, a total of 90 such compact binaries have been detected \citep{gwtc-1, gwtc-2, gwtc-3, gwtc-2.1}. Based on the median mass of the primary and the secondary objects in the binary, the demographics of all the observed LVK events can be broadly categorized into four bins: (i) 83 binary black hole (BBH) \citep{LIGOScientific:2016vlm}, (ii) 3 neutron star - black hole (NSBH) binaries \citep{LIGOScientific:2021qlt}, (iii) 2 binary neutron stars (BNS) \citep{LIGOScientific:2018hze} and (iv) 2 mystery binaries \citep{LIGOScientific:2020zkf}, where in the primary or secondary objects falls in the so-called \textit{lower mass gap} $(2\sim5~M_\odot)$ \citep{Bailyn_1998, _zel_2010, Farr_2011}. Inside this gap, the compact object population is unknown, which will either be a very light stellar black hole or a very heavy neutron star \citep{1980RvMP...52..299T, Zevin2020, LIGOScientific:2020zkf, Dichiara_2021}. Previous studies have explained this mass gap from the inefficiency of core-collapse supernovae and the internal structure of neutron stars \citep{de_S__2022}. A new event from the Fourth Observing Run of LVK (2023-present), GW230529~\citep{2024observation}, with $m1 = 3.6^{+0.8}_{-1.3}~M_\odot$, offers the most conclusive evidence of mystery events in bin (iv). 

Within the 83 LVK BBH events, there are 3 confirmed detections of binaries with the primary and secondary object in the \textit{upper mass gap} $(\sim60-130~M_\odot)$, or more commonly referred in the literature as the Pair-Instability Supernovae (PISN) mass gap \citep{gwtc-1, gwtc-2, gwtc-3, gwtc-2.1}. Stellar evolution and nucleosynthesis theory predict that black holes cannot form \textit{directly} in this entire mass range as the supernova process disrupts the star and leaves no compact remnant \citep{Woosley_2021}. The precise bounds of the PISN mass gap are still uncertain \citep{Fishbach_2020}, especially the lower-bound can span from $[40,~ 60]~M_{\odot}$ \citep{Farmer_2019, Marchant_2020}. The events with the PISN gap are astrophysically interesting as their merger provides direct evidence of intermediate-mass black holes $[10^2, 10^5]~M_{\odot}$ \citep{Greene_2020}. 
The most prominent example of such a binary is GW190521 \citep{LIGOScientific:2020iuh}, in which both $m_1 = 85^{+21}_{-14}~M_{\odot}$ and $m_2 = 66^{+17}_{-18}~M_{\odot}$ are in the PISN mass-gap, and the remnant at $m_\mathrm{f} = 142^{+28}_{-16}~M_{\odot}$ is an intermediate-mass black hole \citep{GW190521astro}. 

Binaries with objects in the lower and upper mass gaps provide a unique opportunity to test new astrophysical formation channels \citep{PhysRevD.95.124046}. For example, hierarchical formation scenario \citep{2017ApJ...840L..24F, Randall_2018, Gerosa_2021}, AGN formation channel \citep{Sigl:2006cg, Bartos_2017, 2021ApJ...908..194T}, and chemically homogeneous evolution \citep{2016MNRAS.458.2634M, 2016MNRAS.460.3545D, Bouffanais_2019} in dense stellar environments \citep{PhysRevLett.121.161103, Randall_2018b, 2020ApJ...903L...5R, antonelli2023classifying, ye2024lowermassgapblackholesdense} can explain populations in both the gaps. While a stable mass transfer during isolated formation channel can produce objects in the lower gap~\citep{Zevin2020, Bavera_2021, van_Son_2022, zhu2024formationlowermassgapblack}.

Comparing the intrinsic properties of events in the lower and upper mass gap (such as masses and spins) with those of the \textit{typical} population also offers insights into their astrophysical origins. For example, the LVK population model for binary black holes shows a characteristic peak just before the expected PISN mass gap~\citep{Farmer_2019, 2021ApJ...913L..23E, 2022ApJ...931..108F, Ye_2022, gwtc-3, 2023PhRvX..13a1048A, Antonini_2023, farah2023thingsbumpnightassessing, chen2024confrontingprimordialblackholes, hernandez2024newbumpnightevidence}.

In this study, we find a novel application of the entropy released from binary black hole mergers for distinguishing between populations in the mass gap from that of overall stellar compact binary population. In Section \ref{sec:method}, we describe our formalism of Merger Entropy Index $(\mathcal{I}_\mathrm{BBH})$ and the phenomenological and population models utilized in this study.

In Section \ref{sec:res}, we report the $\mathcal{I}_\mathrm{BBH}$ distribution for different population models as well as specific LVK events to compare the formation channels and demographics of objects in the upper and lower mass gap.

\begin{figure} [t!]
\centering
\includegraphics[scale=0.4]{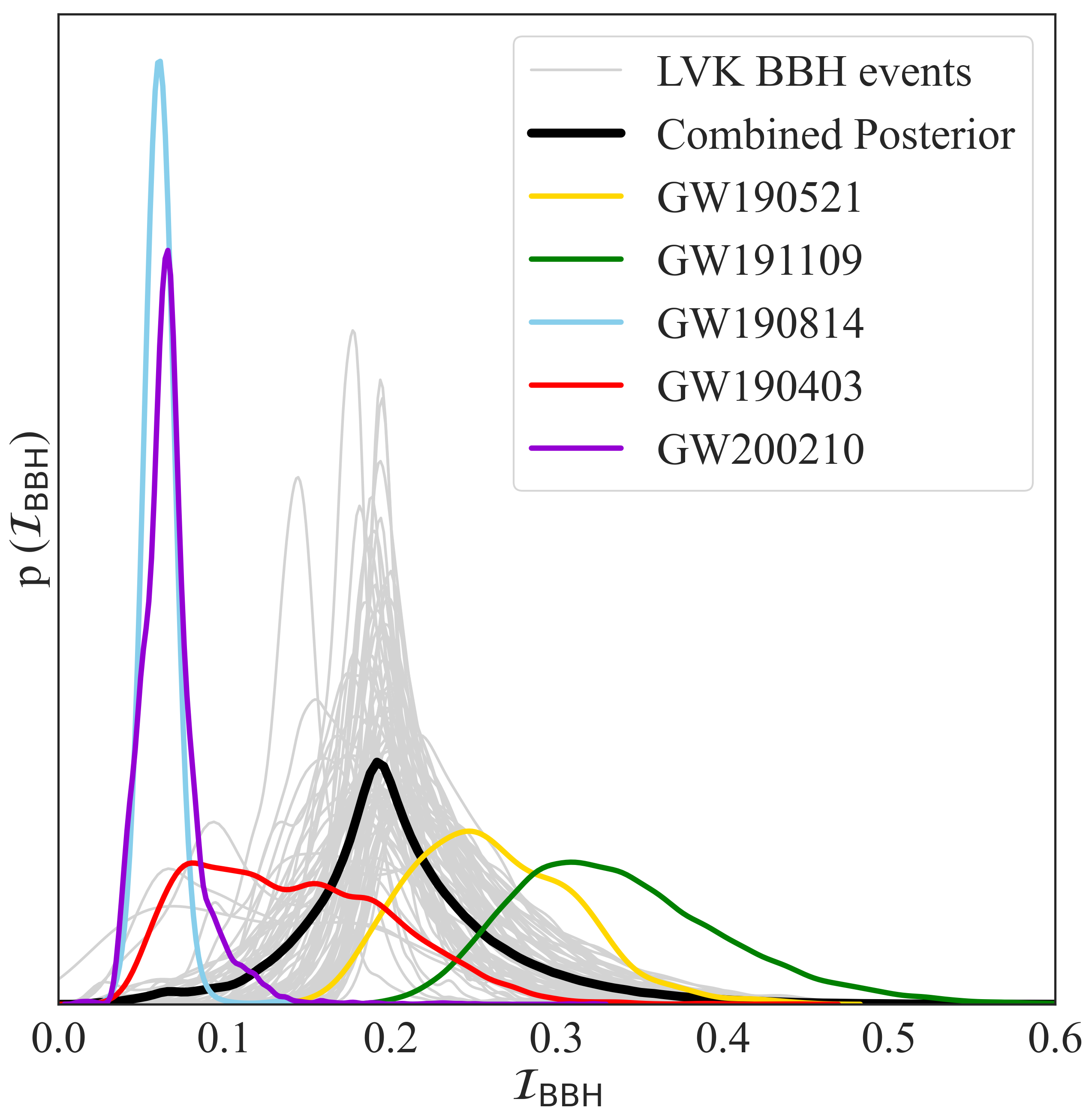}
\caption{
The $\mathcal{I}_\mathrm{BBH}$ posteriors for all mystery and BBH events reported by LVK are shown, with grey lines representing the posteriors for individual events and the black line indicating the combined posterior, intended for visualization purposes only. Certain notable GW events, discussed in detail in Sec. \ref{subsec:resevent}, are highlighted in color. }
\label{ibbh}
\end{figure}

\section{Methods} \label{sec:method}
\subsection{Merger Entropy Index}
\label{subsec:mei}
Based on the Hawking Area Theorem \citep{PhysRevLett.26.1344}, the entropy of a black hole correlates to the surface area of its event horizon. By expressing the Bekenstein-Hawking Formula \citep{PhysRevD.7.2333} in terms of mass (m) and spin ($\chi$), we can compute the entropy of an individual black hole as: 
\begin{equation}
\label{eq:entropy}
\centering
    S = \frac{A}{4l_p^2} = \frac{Ac^2}{4G\hbar^2} = \frac{2\pi G}{\hbar c} m^2 (1+\sqrt{1-\chi^2})
\end{equation}

According to the second law of thermodynamics, mergers between BBHs will increase the entropy of the universe \citep{PhysRevD.7.2333}. Based on this theorem, we introduced a new variable called the Merger Entropy Index ($\mathcal{I}_\mathrm{BBH}$) in an earlier study \citep{hu2021thermodynamics}, which measures the efficiency of entropy transfer during the BBH mergers:

\begin{equation}
\label{eq:deltaentropy}
\centering
    \mathcal{I}_\mathrm{BBH} = \frac{\pi}{9} \frac{\Delta S_{BBH}}{S_1 + S_2}
\end{equation}

where $\Delta S_{BBH} = S_\mathrm{f} - (S_1 + S_2)$ measures the increase in entropy during the merger ($S_\mathrm{f}$ is the entropy of the post-merger black hole, and $S_1,~S_2$ are entropy of the pre-merger black holes). General relativity bounds this parameter between $[0,~1]$. 

For all the GW events reported by LVK, we utilize the parameter estimation results made available in the GW Open Science Center \citep{2021SoftX..1300658A,
 2023ApJS..267...29A}. For BBH events, we use posteriors from SEOBNRv4PHM waveform model \citep{Buonanno_1999, Buonanno_2000, Damour_2008, Ossokine_2020, gadre2022fullyprecessinghighermodesurrogate}, while for NSBH we use the IMRPhenomXPHM posteriors \citep{Khan_2016, Khan_2019, Pratten_2020, Pratten_2021, dambrosio2024testinggravitationalwaveformsgeneral}. The files also include the mass and spin associated with the corresponding remnant black hole, which has been computed using phenomenological fits to NR results \citep{Santamar_a_2010, Hofmann_2016, Healy_2017, Jim_nez_Forteza_2017,  Estell_s_2022}. We input the posteriors of primary, secondary, and remnant black holes in  Eqs. \ref{eq:entropy}-\ref{eq:deltaentropy} and compute the corresponding $\mathcal{I}_\mathrm{BBH}$ posteriors for each LVK event, as highlighted in Fig. \ref{ibbh}.

\subsection{Population Models}
\label{subsec:popmodel}
To test the $\mathcal{I}_\mathrm{BBH}$ formalism for upper and lower mass gap events, we employ six different BBH population models, inspired by astrophysics and detection parameter space. These models are treated as uncorrelated and independent observation channels.

\textit{Uniform Prior} - we adopt the injection file for all LVK observing runs to estimate the searching sensitivity \citep{gwtc-3, 2023PhRvX..13a1048A}. From the injection, we use the parameters of detectable signals without any $p_\mathrm{astro}$ filtering to generate the uniform prior. This model assumes a uniform distribution for spin magnitude $(\chi_{1,2})$ and polar angle $(\theta_{1,2})$ of the spin vectors across the entire range, i.e. $\chi_{1,2} \in [0,~1]$ and $\cos{\theta_{1,2}} \in [-1,~1]$ \citep{Bhagwat_2021}. The distribution of mass ratio $m_2/m_1 \equiv q$ is nearly flat across [0,1], with slightly higher probability on the equal-mass region.

\textit{Isolated BBHs} - in this model, BBHs are expected to have little spin misalignment, so we strictly constrain $\theta_{1,2} \in \{0,~\pi\}$  \citep{2021APS..APRX16009S}. For mass ratio and spin distributions, we assume uniform distributions across the range $q \in [0.5,~1]$, $\chi_{1,2} \in [0,~0.5]$, as shown in \cite{Zevin_2021iso}.

\textit{Dynamical BBHs -} this model includes dynamical capture and hierarchical merger scenario (1G+1G, 1G+2G, 2G+2G) involving second-generational black holes (2G), which are the remnant black holes of 1G mergers \citep{PhysRevLett.123.181101, Kimball_2020, Fragione_2023}. We adopt mass ratio and spin distributions derived in \cite{Kimball_2021} and assume a uniform distribution for polar angle of the spin vectors. Both 1G+1G and 2G+2G populations favor the equal-mass regime, while q distribution of 1G+2G shows a sharp Gaussian distribution peaking at q=0.5. Spin distribution of 1G+1G resembles an exponentially decaying curve with \texttt{MODEL TRUNCGAUSS}, with the highest probability at $\chi=0$. The spin distribution of 2G+2G is a Gaussian distribution with mean of 0.7. $\chi_1, \chi_2$ have different distributions in 1G+2G population, overlapping with the spin distribution of 1G+1G and 2G+2G respectively.

\textit{GWTC-3 PowerLaw + Peak (PP)} - we adopt the PowerLaw + Peak model \citep{Talbot_2018}, which demonstrates the highest Bayes factor in characterizing the Gravitational-Wave Transient Catalog (GWTC) dataset from the Third Observing Run. The most notable feature of this model is the small peak at $\sim 35~M_\odot$ in the primary mass distribution due to the PISN mass gap \citep{Farmer_2019, Farmerconstra_2020, Edelman_2022}. The detailed parameter distributions are described in \cite{2023PhRvX..13a1048A}.

\begin{figure} [t!]
\includegraphics[scale=0.37]{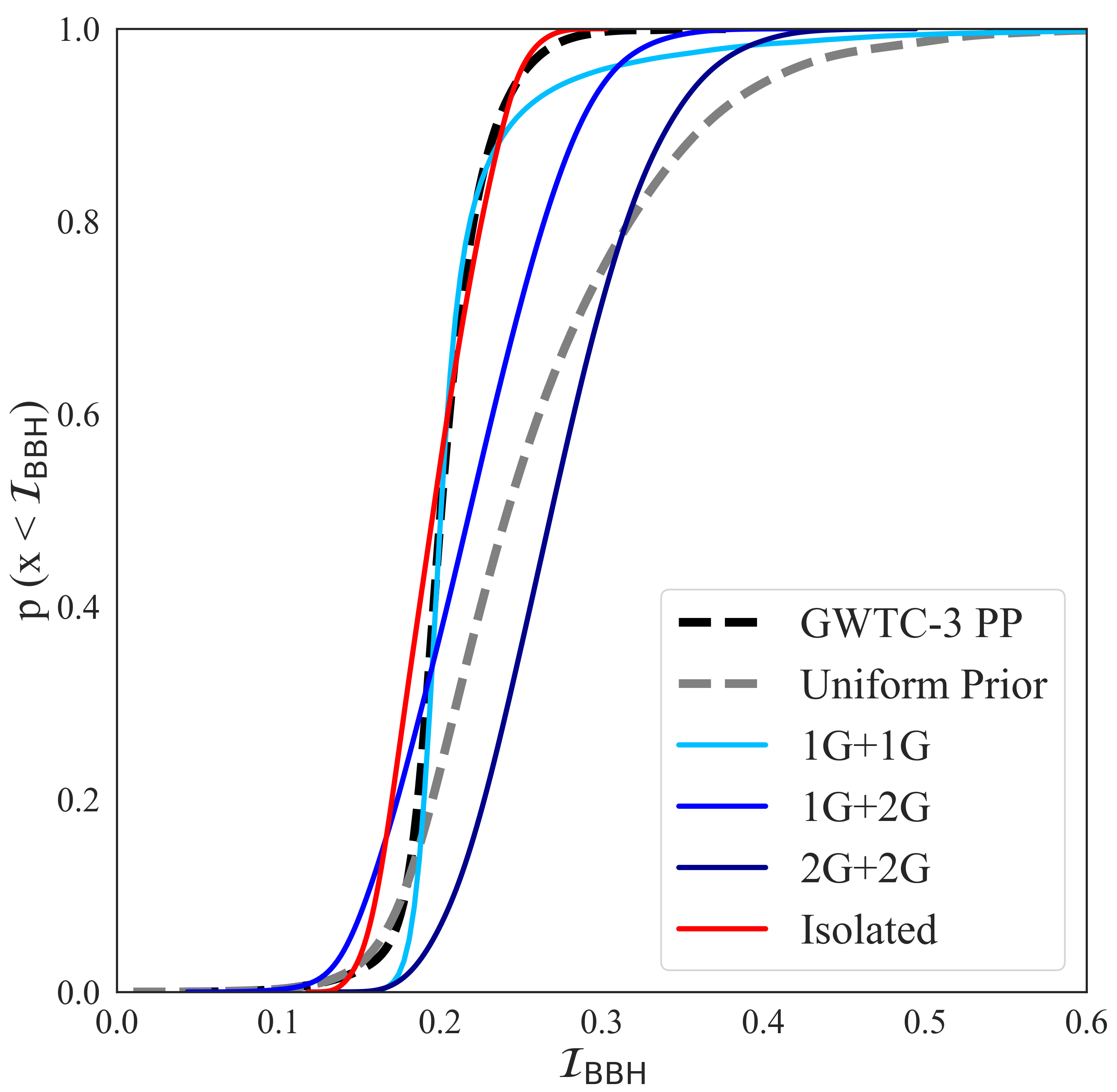}
\caption{CDFs of $\mathcal{I}_\mathrm{BBH}$ posteriors for 6 different population models described in Sec. \ref{subsec:popmodel}. The two dotted lines denote the uniform prior from LVK injection and merging populations of GWTC-3 as reference. Blue curves show $\mathcal{I}_\mathrm{BBH}$ of black hole populations with different generations in dynamical channels, and the red curve shows $\mathcal{I}_\mathrm{BBH}$ of black holes in isolated channels.}
\label{imbh}
\end{figure}

For each population model, we randomly sample $i=10^4$ binaries with inspiral parameters $\Lambda^\mathrm{ins}_i: \{ q,~\chi_{1,2},~\theta_{1,2} \}$. The bounds on $\Lambda^\mathrm{ins}_i$ are dictated by the population model. To obtain the corresponding parameters of the remnant black hole, $\Lambda^\mathrm{rem}_i: \{ m_\mathrm{f},~\chi_\mathrm{f} \}$, we utilize the phenomenological fit \texttt{NRSur7dq4Remnant} \citep{Blackman_2017, Boschini_2023} from the \textit{SurfinBH} package \citep{soft09007V, Varma_2019, zertuche2024highprecisionringdownsurrogatemodel}. For each $\Lambda_i$, we compute $\mathcal{I}_{\mathrm{BBH},i}$ using Eq. \ref{eq:entropy} - \ref{eq:deltaentropy}. The high sampling rate provides sufficient accuracy to compare the posteriors of $\mathcal{I}_{\mathrm{BBH}}$ between different population models and to test them with those from LVK events. The probability distribution derived from the random sampling for each population model is shown in Fig. \ref{imbh}.

\subsection{Comparison between Models and Events}
\label{subsec:ks}

The catalog of GW events observed by the LVK detectors is affected by the selection effects \citep{Singer_2014, LIGOScientific:2016vlm, 2017ApJ...835...31C}. This introduces a bias when constructing the $\mathcal{I}_\mathrm{BBH}$ posteriors. To correct for the selection effect, we utilize the formalism described in \cite{Doctor_2021} by introducing a weight factor for each $\Lambda_{i}$:

\begin{equation}
\centering
\mathrm{w}_{i} = \frac{p(\Lambda_{i}|z_\mathrm{det})}{p(\Lambda_{i}|z_\mathrm{ref})}
\end{equation}

We obtain $p(\Lambda_{i}|z_\mathrm{det}, z_\mathrm{ref})$ from the injection file of combined three LVK observing runs. To compute $p(\Lambda_{i}|z_\mathrm{det})$, we adopt the \texttt{PYCBC BBH} detection pipeline and compute $p_\mathrm{astro}$ for each inspiral parameter \citep{Tiwari_2018, Farr_2019, 2023PhRvX..13a1048A}. We only take detectable $\Lambda_{i}$ with $p_\mathrm{astro} \geq 0.5$. We then apply $\mathrm{w}_{i}$ to the post-merger parameters  $\Lambda^\mathrm{rem}_i$ derived from numerical relativity in Sec. \ref{subsec:popmodel}. All population models used have already accounted for selection effect. Since the selection effect is relevant when comparing population instead of individual event \citep{Thrane_2019}, we only correct when grouping LVK events together.

After correcting for the selection effect, we perform Kolmogorov–Smirnov (KS) tests \citep{Plowman_2011} among population models as well as their comparison with individual LVK events. The KS tests compute $\mathcal{D}$ statistics, which is a measure of the inherent difference between the two distributions.

In this study, we quote \textit{favored} population model for individual and group of LVK events based on $(1-\mathcal{D})$. These values have been quoted in Table \ref{tab:table1} and Fig. \ref{ks}.

\section{Results \& Discussions}
\label{sec:res}

\subsection{$\mathcal{I}_\mathrm{BBH}$ of LVK BBH Events}
\label{subsec:resevent}
In Fig. \ref{ibbh}, we showcase the $\mathcal{I}_\mathrm{BBH}$ posteriors for 85 LVK events that broadly qualify as BBH systems. For further investigation, we group these BBH events into four categories based on the median value of the component masses in their cosmologically corrected source frame: (i) 74 stellar BBH events in which $m_{1,2} \in [5,~60]~M_\odot$; (ii) 3 PISN-PISN events with $m_1, m_2 > 60~M_\odot$, wherein both the primary and secondary are in upper gap $(60-130~M_\odot)$; (iii) 6 PISN BH-Stellar BH events with $m_1 > 60~M_\odot$, wherein only the primary object is in upper gap, and (iv) 2 stellar BH-mystery BH events, wherein the secondary object is in the lower gap $(2-5~M_\odot)$. Table \ref{tab:tabevent} highlights the GW events in the special categories (ii)-(iv) and their corresponding $\mathcal{I}_\mathrm{BBH}$ values.

\begin{table} [t!]
\caption{Highlighted special GW events in this work, which are also the categories we classify GW events. Stellar black holes are [5,~60]~$M_\odot$ (74 in total). $m_1,m_2$ refers to the median mass measurements in the source frame. $\mathcal{I}_\mathrm{BBH}$ values are present with 90\% confidence interval.}

\begin{ruledtabular}
\begin{tabular}{c|c|c}
Event Type & Name & $\mathcal{I}_\mathrm{BBH}$\\
\hline
$m_1,m_2\geq60~M_\odot$ & GW190426 & $0.21^{+0.11}_{-0.08}$  \\
\textit{(ii: Both in upper gap)} & GW190521 & $0.26^{+0.09}_{-0.07}$ \\
 & GW200220 & $0.22^{+0.13}_{-0.06}$ \\
\hline
$m_1\geq60~M_\odot$ & GW190403 & $0.14^{+0.11}_{-0.08}$ \\
\textit{(iii: Primary in upper gap)} & GW190519 & $0.19^{+0.08}_{-0.04}$ \\
 & GW190602 & $0.20^{+0.11}_{-0.06}$ \\
 & GW190706 & $0.18^{+0.08}_{-0.05}$ \\
 & GW190929 & $0.18^{+0.12}_{-0.06}$ \\
 & GW191109 & $0.33^{+0.13}_{-0.08}$ \\
\hline
$m_2 \in [2,5]~M_\odot$ & GW190814 & $0.06^{+0.01}_{-0.01}$\\
\textit{(iv: Secondary in lower gap)} & GW200210 & $0.06^{+0.03}_{-0.02}$\\
\hline 
$m_2 \in [1,2]~M_\odot$ & GW191219 & $0.02^{+0.01}_{-0.00}$\\
\textit{(Stellar BH + NS )} & GW200105 & $0.10^{+0.03}_{-0.01}$\\
 & GW200115 & $0.12^{+0.22}_{-0.04}$\\
\hline
$m_1 \in [2,5]~M_\odot$ & GW230529 & $0.18^{+0.07}_{-0.08}$\\
\textit{(Primary in lower gap + NS)} & & \\
\end{tabular}
\label{tab:tabevent}
\end{ruledtabular}
\end{table}

\begin{table*} [t!]
\caption{The similarity of $\mathcal{I}_\mathrm{BBH}$ between different BBH population models. The lower the value, the more distinction between the two models.}
\begin{ruledtabular}
\begin{tabular}{c|cccccc}
  & Uniform & 1G+1G & 1G+2G & 2G+2G & Isolated & GWTC-3 PP\\
\hline
Uniform & / & 0.55 & 0.81 & 0.76 & 0.60 & 0.57\\
1G+1G & 0.55 & / & 0.68 & 0.34 & 0.67 & 0.85 \\
1G+2G & 0.81 & 0.68 & / & 0.64 & 0.73 & 0.70\\
2G+2G & 0.76 & 0.34 & 0.64 & / & 0.37 & 0.34\\
Isolated & 0.60 & 0.67 & 0.73 & 0.37 & / & 0.79\\
GWTC-3 PP & 0.57 & 0.85 & 0.70 & 0.34 & 0.79 & /\\
\end{tabular}
\label{tab:table1}
\end{ruledtabular}
\end{table*}

For the combined 85 LVK BBH events, we find that $\mathcal{I}_\mathrm{BBH} = 0.20^{+0.12}_{-0.08}$ with 90\% confidence intervals. This is consistent with our expectation for the astrophysical prior of $\mathcal{I}_\mathrm{BBH} \leq 0.3$ derived in the earlier work \citep{hu2021thermodynamics}. For categories describing (i) stellar BBH, we find  $\mathcal{I}_\mathrm{BBH} = 0.20^{+0.11}_{-0.08}$; (ii) PISN-PISN BBH, we find  $\mathcal{I}_\mathrm{BBH} = 0.21^{+0.11}_{-0.08}$; (iii) PISN BH-Stellar BH, we find  $\mathcal{I}_\mathrm{BBH} = 0.20^{+0.16}_{-0.07}$; (iv) stellar BH-mystery BH, we find  $\mathcal{I}_\mathrm{BBH} = 0.06^{+0.02}_{-0.01}$. For each of these cases, the combined posterior of  $\mathcal{I}_\mathrm{BBH}$ is corrected for the selection effects as described in Sec. \ref{subsec:ks}.

Among all LVK events, GW191109 shows the largest $\mathcal{I}_\mathrm{BBH}$ = $0.33^{+0.13}_{-0.08}$, followed by GW190521 $\mathcal{I}_\mathrm{BBH}$ = $0.26^{+0.09}_{-0.07}$. Both these events belong to the high-mass population (categories ii, iii), with primary masses of $65^{+11}_{-11}$ and $85^{+21}_{-14}~M_\odot$ respectively. Other most massive LVK BBHs, GW190426 (from GWTC-2.1 \citep{gwtc-2.1}) and GW200220, have slightly higher $\mathcal{I}_\mathrm{BBH}$ than the combined stellar BBH events.

As shown in our earlier work \citep{hu2021thermodynamics},  $\mathcal{I}_\mathrm{BBH}$ has a strong dependence on the orientations of black hole spins. For example, a binary with two highly spinning black holes can either result in a remnant black hole with a moderate spin or maximal spin depending on the spin evolution during the inspiral. Yet, the first case would result in a significant increase in entropy than the latter. $\mathcal{I}_\mathrm{BBH}$ is particularly sensitive to the effective inspiral spin parameter $\chi_\mathrm{eff}$ when the binaries have near to moderate mass-ratios. A negative (positive) value of $\chi_\mathrm{eff}$ generally results in a higher (lower) entropy configuration and thus a larger (smaller) value for $\mathcal{I}_\mathrm{BBH}$. As we demonstrate in Sec. \ref{subsec:resLVK}, this feature of $\mathcal{I}_\mathrm{BBH}$ makes for a powerful probe to test astrophysical formation channels for a GW event. 

In the case of GW191109, the higher value of $\mathcal{I}_\mathrm{BBH}$ can be mainly attributed to significant anti-alignment of black hole spins with respect to the orbital angular momentum ($\chi_\mathrm{eff} = -0.29^{+0.42}_{-0.31}$). The spin configuration of GW191109 provides hints to dynamical formation scenario \citep{shaw2022, antonelli2023classifying, zhang2023likely}. However, GW191109 suffers from data quality issues and glitch \citep{theligoscientificcollaboration2021tests, Wang_2022, Udall:2024ovp}, which affects the parameter estimation including negative $\chi_\mathrm{eff}$ and can be the origin of an inconsistent $\mathcal{I}_\mathrm{BBH}$ in comparison to other LVK events.

On the other hand, GW190403 has the lowest $\mathcal{I}_\mathrm{BBH} = 0.14^{+0.11}_{-0.08}$ among all the 9 events in the high-mass population. This is consistent with expectations since the GW190403 has highest recorded spin-orbit alignment among all LVK events ($\chi_\mathrm{eff} = 0.68^{+0.16}_{-0.43}$), thus a stronger preference for smaller $\mathcal{I}_\mathrm{BBH}$. The large positive $\chi_\mathrm{eff}$ of GW190403 indicates significant alignment between spin and orbit, providing hints to isolated formation scenario \citep{Belczynski_2020, 2021ApJ...921L...2O, Bavera_2020, Qin_2022, gwtc-2.1}.

Furthermore, GW190403 is the most asymmetric binary in the high-mass population $(q={0.25}_{-0.11}^{+0.54})$. Highly asymmetric BBH has the least entropy transfer and therefore drives $\mathcal{I}_\mathrm{BBH}$ towards its lower bound. As seen in Fig. \ref{ibbh}, GW190403 also shows the largest standard deviation in $\mathcal{I}_\mathrm{BBH}$ distribution among all the LVK events. This event has a low SNR of ${7.6}^{+0.6}_{-1.1}$ \citep{gwtc-2.1}, resulting in a much larger uncertainty in the inspiral parameters, and subsequently broader $\mathcal{I}_\mathrm{BBH}$ posterior. Yet, the GW190403 does exhibit multiple peaks in its $\mathcal{I}_\mathrm{BBH}$ distribution, which is characteristic of moderate mass-ratio binary being impacted by its spin orientations. 

The link between asymmetric BBH events and lower $\mathcal{I}_\mathrm{BBH}$ is further illustrated by GW190412 \citep{LIGOScientific:2020stg}, which shows a relatively smaller $\mathcal{I}_\mathrm{BBH} \approx 0.14 $ in comparison to the combined posterior of stellar BBHs peaking at $\mathcal{I}_\mathrm{BBH} \approx 0.20$. The unequal mass ratio is not common in isolated pathways, which points to hierarchical merger origin for these two events \citep{Banerjee_2022, Ford_2022}.

The formalism of $\mathcal{I}_\mathrm{BBH}$ is deemed unsuitable for BNS or NSBH events due to the uncertainty in the entropy calculation for their pre- and post-merger objects. However, the formalism can be applied to GW events with secondary objects in the lower mass gap (category-iv), such as GW190814 and GW200210 \citep{LIGOScientific:2020zkf, gwtc-2, gwtc-3}, by treating them as BBH. As shown in Fig. \ref{ibbh}, both these events in the lower mass gap display sharp spikes around $\mathcal{I}_\mathrm{BBH} \sim 0.06$. Since the mass-ratio of these binaries is $q \sim 0.1$, their spin orientations $(\chi_\mathrm{eff}\sim0)$ have no substantial impact on $\mathcal{I}_\mathrm{BBH}$ unless the black holes are highly anti-aligned \citep{hu2021thermodynamics}. Since the secondary mass for GW190814 is measured and constrained more precisely than GW200210, their $\mathcal{I}_\mathrm{BBH}$ posterior is narrower.

\subsection{$\mathcal{I}_\mathrm{BBH}$ of Population Models}
\label{subsec:respop}

In Fig. \ref{imbh}, we highlight the distribution of $\mathcal{I}_\mathrm{BBH}$ for all six population models described in Sec. \ref{subsec:popmodel}. Note that across all the models, $\mathcal{I}_\mathrm{BBH} \gtrsim 0.1$. This lower bound is dictated by the fact that all six models predict near-equal to moderate mass-ratio BBH systems. On the other bound, none of the models predict $\mathcal{I}_\mathrm{BBH} \gtrsim 0.6$. This upper bound results from our assumption that binaries in all of the models are quasi-circular, which is a simplification since the dynamical channel could have some residual eccentricity event in the LVK frequency range~\citep{Romero_Shaw_2020, Romero_Shaw_2022}. As discussed previously in \cite{hu2021thermodynamics}, highly eccentric binaries with spins are more efficient than circular binaries for entropy transfer, thus resulting in a larger $\mathcal{I}_\mathrm{BBH}$.

The population model 2G+2G results in the largest value of $\mathcal{I}_\mathrm{BBH} = 0.27^{+0.09}_{-0.08}$, while the isolated formation model has the lowest value of $\mathcal{I}_\mathrm{BBH} = 0.20^{+0.05}_{-0.04}$. We find that value of $\mathcal{I}_\mathrm{BBH}$ decreases with the generations of hierarchical merger, with 2G+2G being the highest, followed by 1G+2G ($\mathcal{I}_\mathrm{BBH} = 0.22^{+0.08}_{-0.07}$) and then 1G+1G $(\mathcal{I}_\mathrm{BBH} = 0.20^{+0.09}_{-0.01})$. In comparison, the signature GWTC-3 Powerlaw+Peak (PP) model shows $\mathcal{I}_\mathrm{BBH} = 0.20^{+0.05}_{-0.03}$, which, as expected, is consistent with the combined posterior of all LVK events discussed in Sec. \ref{subsec:resevent}. When considering the uniform prior for BBH,  $\mathcal{I}_\mathrm{BBH} $ shows the broadest distribution from $ 0.24^{+0.16}_{-0.07}$. 

We gain further insights into the distinction between population models by conducting KS test of $\mathcal{I}_\mathrm{BBH}$ distribution. We summarize the KS test results in Table \ref{tab:table1}. We find that the dynamical capture 1G+1G and GWTC-3 PP has the highest similarity of 85\% between their $\mathcal{I}_\mathrm{BBH}$ distributions. Closely followed is the similarity of 79\% between isolated formation model and GWTC-3 PP. Both 1G+1G and isolated formation have similar mass ratio distribution that favors the equal mass region \citep{Giacobbo_2018, Bouffanais_2019, 2020ApJ...903L...5R}. We also note that the hierarchical merger scenario of 2G+2G is the least similar (34\%) with the  $\mathcal{I}_\mathrm{BBH}$ distribution of GWTC-3 PP. These comparisons are well within the expectations \citep{Zevin_2021iso} and demonstrate that $\mathcal{I}_\mathrm{BBH}$ could be an independent probe for the underlying population of the observed BBH events.  

Since $\mathcal{I}_\mathrm{BBH}$ is mass invariant by definition, it will not recognize features such as the peak in GWTC-3 PP model before the PISN mass-gap or that the 2G+2G scenario is applicable only to events inside the PISN mass-gap. Yet, the index is sensitive to the underlying spin and mass-ratio differences between these models. Therefore, we notice that $\mathcal{I}_\mathrm{BBH}$ distribution of 1G+1G (lower mass BBH) has the least similarity with 2G+2G (34\%). In the latter model, the spin distribution peaks $\sim$ 0.7 as the binary transports the angular momentum from the earlier first-generation merger to the remnant black hole. Similarly, the $\mathcal{I}_\mathrm{BBH}$ distribution of 1G+1G favors the uniform prior only with 55\%, which is a reflection of the mass-ratio being near equal for dynamical capture of the first generation mergers \citep{Kimball_2021}. The low similarity of $\mathcal{I}_\mathrm{BBH}$ for 2G+2G model in comparison to all other models enables us to distinguish within dynamical channels as well as isolated formation channels.    

On the other hand, 1G+2G has the most generalizable $\mathcal{I}_\mathrm{BBH}$ distribution among all population models in this study. As shown in Table \ref{tab:table1}, the similarity of 1G+2G model ranges from $81\%$ for uniform prior, $73\%$ for isolated binaries to the lowest of $64\%$ with 2G+2G. While the 1G+2G model is referring to events with primary black hole in the upper gap, the underlying spin and mass-ratio distribution has much larger overlap with other channels that are suitable only for lower or higher masses. For example, as described in Sec. \ref{subsec:popmodel}, isolated has a uniform $q$ distribution within [0.5,~1], 2G+2G population is mostly composed of $q\sim1$, while 1G+2G prefer $q = 0.5$. In the next section, we discuss that 1G+2G model is indeed the most consistent dynamical formation channel with LVK events in categories-ii,iii.

\begin{figure*} [hbt!]
\centering
\includegraphics[scale=0.25]{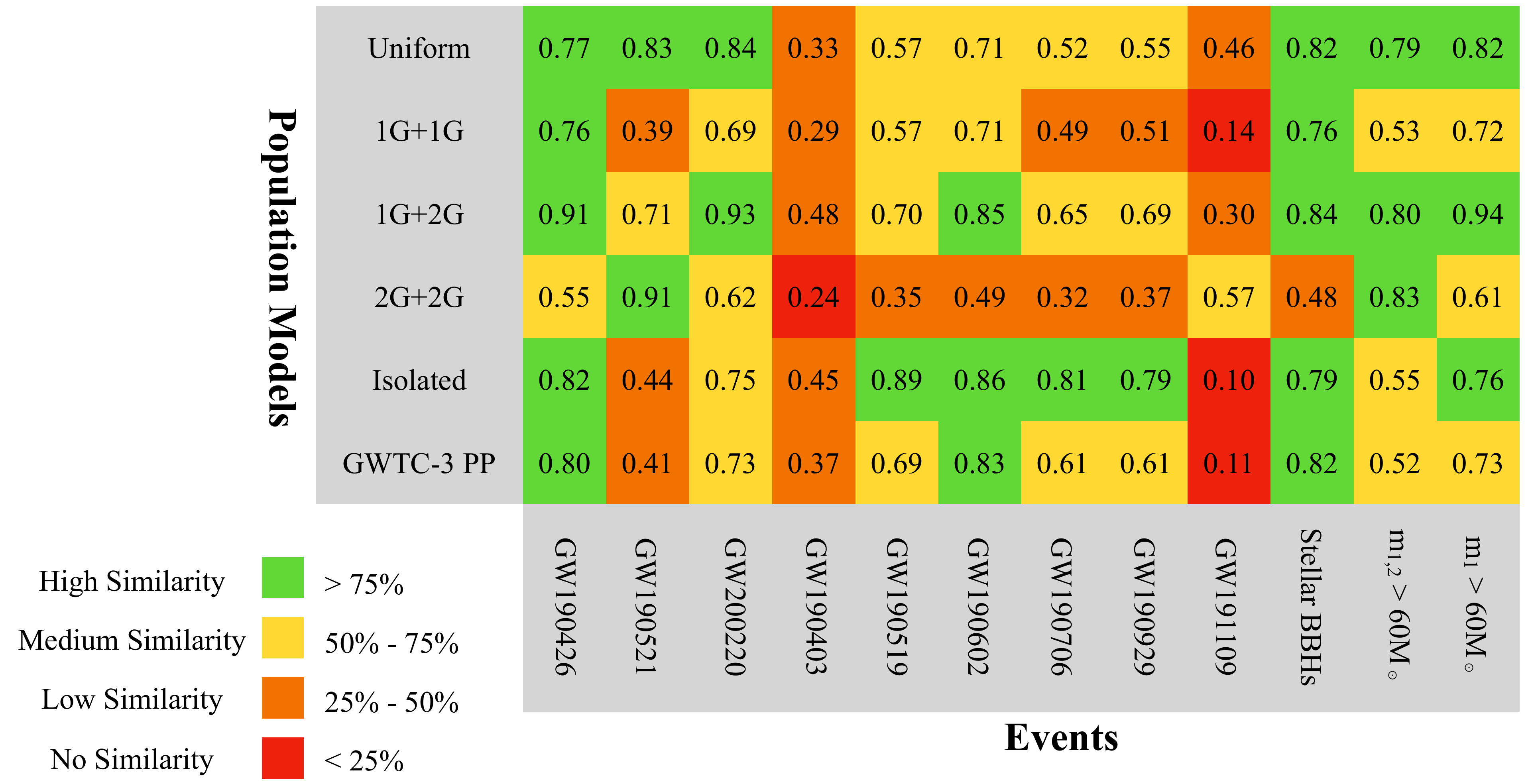}
\caption{The similarity of $\mathcal{I}_\mathrm{BBH}$ posteriors between population model and individual GW event / population reported by LVK. The color scheme visualizes the similarity with 4 categories, with the number denoted as the percentage of similarity in $\mathcal{I}_\mathrm{BBH}$. The individual events come from all the PISN-PISN \& PISN-Stellar BH binaries listed in Table \ref{tab:tabevent}. In the last three columns, we also group stellar BBHs (74), PISN-PISN BBHs (3), and PISN BH-Stellar BH binaries (6) together after correcting for selection effect in the detector frame.}
\label{ks}
\end{figure*}

\subsection{Comparing Population Models with LVK Upper Mass Gap Events}
\label{subsec:resLVK}

Fig. \ref{ks} shows the results of KS convergence tests between $\mathcal{I}_\mathrm{BBH}$ posteriors of 6 population models described in Sec. \ref{subsec:popmodel} and individual LVK events from category ii, iii in Table \ref{tab:tabevent}. The consistency between BBH groups and uniform prior serves as a baseline for comparison. For categories describing (i) stellar BBH: $\mathcal{I}_\mathrm{BBH}$ shows the highest consistency with 1G+1G (76\%) isolated models (79\%), while only 48\% consistency is observed with the 2G+2G population; (ii) PISN-PISN BBH: $\mathcal{I}_\mathrm{BBH}$ favors 2G+2G (83\%); (iii) PISN BH-Stellar BH: $\mathcal{I}_\mathrm{BBH}$ favors 1G+2G (94\%). 
$\mathcal{I}_\mathrm{BBH}$ of GWTC-3 PP model shows an 82\% match with category i, but only a 52\% match with category ii.  The high spin typical of the 2G+2G population contributes to the lower consistency in $\mathcal{I}_\mathrm{BBH}$ between the PISN-PISN BBH and the GWTC-3 population. Stellar BBH (category i) also make up the majority of the GWTC-3 population \citep{2023PhRvX..13a1048A}.

GW190521, the second-most massive BBH event reported to date, with both black holes residing in the PISN mass gap, has $\mathcal{I}_\mathrm{BBH} = 0.26^{+0.09}_{-0.07}$, which shows a 91\% match with 2G+2G model. In contrast, $\mathcal{I}_\mathrm{BBH}$ of this event has lower consistency to other models, especially for 1G+1G by only 39\% similarity. $\mathcal{I}_\mathrm{BBH}$ distribution favors the 2G+2G model, suggesting that the high progenitor masses likely result from hierarchical mergers. This conclusion matches with results from many recent literature \cite{Romero_Shaw_2020, LIGOScientific:2020iuh, Graham_2020, Estelles:2021jnz, Holgado_2021, Gamba_2022, morton2023gw190521binaryblackhole}. Additionally, $\mathcal{I}_\mathrm{BBH}$ of GW190521 has low consistency (41\%) with GWTC-3 populations, indicating that a flatter PP model may be required to fit these high-mass events \citep{Kimball_2021}.

The most massive BBH detected so far in GW190426 has $\mathcal{I}_\mathrm{BBH} = 0.21^{+0.11}_{-0.08}$, which is consistent with $\mathcal{I}_\mathrm{BBH}$ posteriors of GWTC-3 population by 80\%. Unlike GW190521, $\mathcal{I}_\mathrm{BBH}$ of this event matches with 2G+2G the least (55\%), while shows higher consistency with 1G+2G (91\%) and isolated (82\%) models. The primary black hole, with a mass of $106^{+45}_{-24}~{M}_{\odot}$, can reasonably be second-generational. However, it is unlikely that the secondary black hole would reach $76^{+26}_{-36}~{M}_{\odot}$ solely through dynamical capture mechanisms. With our 1G+2G model described in Sec. \ref{subsec:popmodel}, P$(m_2 = 76~{M}_{\odot}) \sim 0$, suggesting that GW190426 is unlikely to be a 1G+2G binary despite its high consistency in $\mathcal{I}_\mathrm{BBH}$. GW190426 shows no orbital eccentricity \citep{Romero_Shaw_2022}, which is a signature of dynamical formation channel. This lack of eccentricity, combined with the 82\% $\mathcal{I}_\mathrm{BBH}$ consistency with the isolated model, suggests that this massive BBH may evolve from isolated formation channels.

GW200220 shows the highest similarity (93\%) of $\mathcal{I}_\mathrm{BBH}$ with a population model (1G+2G). The secondary black hole $m_2 = 61^{+21}_{-25}~{M}_{\odot}$ lies near the boundary of the PISN mass gap ($60~{M}_{\odot}$) by our definition. However, the exact bound of the PISN mass gap remains uncertain and may extend to 65~$M_\odot$ \citep{Woosley_2021}. When defining PISN mass gap as $[65,~130]~M_\odot$, the mass distribution of GW200220 matches with 1G+2G binary. The secondary black hole may be first-generational given P($m_2 \leq$ 65~$M_\odot$) $\geq 60\%$. In contrast to GW190426, GW200220 shows explicitly higher consistency in $\mathcal{I}_\mathrm{BBH}$ with 1G+2G (93\%) than isolated models (75\%), so $\mathcal{I}_\mathrm{BBH}$ disfavors the isolated formation channel.

On the other hand, GW191109 has the least similarity across all population models. The $\mathcal{I}_\mathrm{BBH}$ posterior for this event is only 10\% consistent with isolated BBHs and 14\% consistent with 1G+1G binaries, but shows a relatively higher 57\% match with the 2G+2G population, supporting the hierarchical merger model for the massive primary black hole and spin configuration. Since GWTC-3 PP model is derived with a strong bias towards $\chi_\mathrm{eff} \sim 0$ \citep{2023PhRvX..13a1048A}, $\mathcal{I}_\mathrm{BBH}$ between GW191109 and GWTC-3 PP only matches by 11\%. As suggested in Sec. \ref{subsec:resevent} and Table \ref{tab:tabevent}, this event has the largest $\mathcal{I}_\mathrm{BBH}$ due to significant spin-orbit misalignment. Since this event does not show high similarity (green category) with any population model, we provide an independent confirmation that GW191109 may be affected by a glitch.

For all the six events in category iii, we note their $\mathcal{I}_\mathrm{BBH}$ favors 1G+2G model more than 1G+1G model. This is consistent with the simulations from \cite{Kimball_2021}, which show PISN BH-Stellar BH binaries have mass ratio distributions consistent with 1G+2G population. The $\mathcal{I}_\mathrm{BBH}$ distribution of GW190602 is 85\% favored towards 1G+2G and about the same as the isolated channel, making it challenging to differentiate between them. However, the remaining events, GW190519, GW190706, and GW190929, show a slight preference for the isolated formation channel. Due to the low SNR and broader $\mathcal{I}_\mathrm{BBH}$ posterior discussed in Sec. \ref{subsec:resevent}, the $\mathcal{I}_\mathrm{BBH}$ formalism of GW190403 does not favor any of six population model with medium similarity.

\subsection{Comparing Population Models with LVK Lower Mass Gap Events}
\label{subsec:nsbh}

\begin{figure}[t!]
\centering
\includegraphics[scale=0.15]{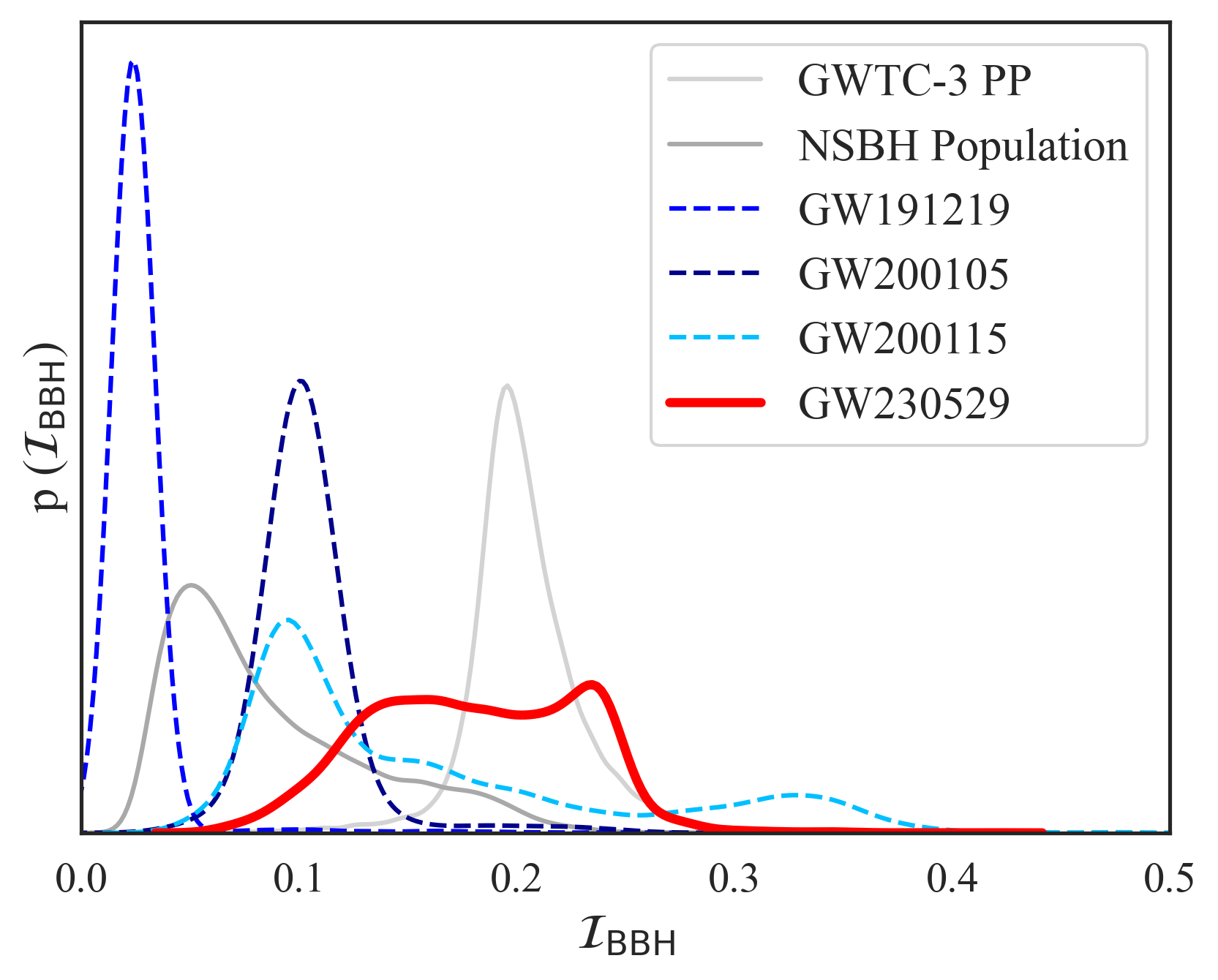} 
\caption{$\mathcal{I}_\mathrm{BBH}$ posteriors of GW230529 with three confirmed NSBH events (blue dotted), GWTC-3 PP model, and NSBH population (grey solid) as reference.}
\label{0529}
\end{figure}

In April 2024, LVK announced GW230529 from O4 \citep{2024observation}, a potential BNS coalescence or NSBH merger event. The primary compact object has a mass of $3.6^{+0.8}_{-1.3}M_\odot$, which sits in the lower mass gap with 99\% confidence interval. The primary object is too heavy to fit into our current understanding of neutron star mass distribution \citep{2012ApJ...757...55O, Kiziltan_2013, Landry_2021} and the neutron star equation of state \citep{Godzieba_2021}. The secondary object ($m_2 = 1.4^{+0.6}_{-1.2}M_\odot$) is likely a neutron star. 

To understand the demographics of GW230529, we apply the $\mathcal{I}_\mathrm{BBH}$ formalism to the available parameter estimation samples from the IMRPhenomXHM model \citep{PhysRevD.103.104056}. This model does not incorporate tidal disruptions expected \citep{zhu2024formation}, but instead assumes coalescence between light black holes, thus permitting us to compute the merger entropy. The  $\mathcal{I}_\mathrm{BBH}$ distribution of GW230529 is shown in Fig. \ref{0529}. Alongside, we have included $\mathcal{I}_\mathrm{BBH}$ distribution from the three known NSBH events  GW191219, GW200105, and GW200115. For these NSBH events, we have the samples from IMRPhenomXHM model to compute $\mathcal{I}_\mathrm{BBH}$. 

When comparing with individual LVK NSBH events, we find that $\mathcal{I}_\mathrm{BBH}$ of GW230529 shows the highest similarity with GW200115 (61\%), which is a confirmed NSBH event with $q = 0.26^{+0.35}_{-0.10}$. In contrast, GW230529  favors the other two asymmetric NSBH events, GW191219 and GW200105, only with 1.5\% and 17\% respectively. 

Furthermore, we derive the $\mathcal{I}_\mathrm{BBH}$ distribution for an NSBH population model described in \cite{Biscoveanu_2022}. This model features the parameter distributions for binaries with secondary compact object $\lesssim 5~M_\odot$. Here, the mass ratio distribution peaks at $q \sim 0.1$. The spin distribution of neutron stars is assumed to be uniform between 0 and 0.7. The $\mathcal{I}_\mathrm{BBH}$ posterior of NSBH population model peaks $\sim 0.06$. We find that the $\mathcal{I}_\mathrm{BBH}$ of GW230529 favors the GWTC-3 PP model (60\%) over the NSBH population model (33\%).

\section{Conclusion}
In this work, we demonstrate a novel application of the Merger Entropy Index ($\mathcal{I}_\mathrm{BBH}$) formalism for probing the astrophysical formation channels of compact objects. We apply this formalism to all the confirmed GW events reported by the LVK collaboration so far, except the two BNS events where it is not applicable. We also derive the $\mathcal{I}_\mathrm{BBH}$ distribution for the six distinct astrophysical population models, including the signature PowerLaw + Peak model from Gravitational-wave Transient Catalogs and hierarchical capture models. We find that the $\mathcal{I}_\mathrm{BBH}$ formalism favors the second-generation merger scenarios for the majority of LVK events in the upper mass gap, while it does not rule out isolated formation scenario for the ones in the stellar mass range. Since the $\mathcal{I}_\mathrm{BBH}$ formalism is mass invariant, we apply it to events in the lower mass gap and compare them with the NSBH binary population. Our work provides a new framework to categorize compact binary events in this data-rich era of gravitational-wave astronomy.

\section{Acknowledgment}
We thank Chase Kimball, Darsan Bellie, and Sylvia Biscoveanu for insightful discussions. We acknowledge the support from Littlejohn Fellowship and Vanderbilt University Summer Research program. This material is based upon work supported by NSF’s LIGO Laboratory which is a major facility fully funded by the National Science Foundation. This research has made use of data, software and/or web tools obtained from the Gravitational Wave Open Science Center (\url{https://www.gw-openscience.org/}), a service of LIGO Laboratory, the LIGO Scientific Collaboration and the Virgo Collaboration. LIGO Laboratory and Advanced LIGO are funded by the United States National Science Foundation (NSF) as well as the Science and Technology Facilities Council (STFC) of the United Kingdom, the Max-Planck-Society (MPS), and the State of Niedersachsen/Germany for support of the construction of Advanced LIGO and construction and operation of the GEO600 detector. Additional support for Advanced LIGO was provided by the Australian Research Council. Virgo is funded, through the European Gravitational Observatory (EGO), by the French Centre National de Recherche Scientifique (CNRS), the Italian Istituto Nazionale di Fisica Nucleare (INFN) and the Dutch Nikhef, with contributions by institutions from Belgium, Germany, Greece, Hungary, Ireland, Japan, Monaco, Poland, Portugal, Spain. 

\bibliography{sample631}{}
\bibliographystyle{aasjournal}

\end{document}